\documentclass{arxiv}

\usepackage{siunitx}
\title{Icy worlds: Moons and Dwarf Planets}

\author[1]{Jun Kimura}
\affil[1]{Department of Earth and Space Science, The University of Osaka, Toyonaka, Japan}

\date{\today}

\begin{document}
\maketitle

\begin{abstract}
In the outer solar system beyond Jupiter, water ice is a dominant component of planetary bodies, and most solid objects in this region are classified as icy bodies. Icy bodies display a remarkable diversity of geological, geophysical, and atmospheric processes, which differ fundamentally from those of the rocky terrestrial planets. Evidence from past and ongoing spacecraft missions has revealed subsurface oceans, cryovolcanic activity, and tenuous but persistent atmospheres, showing that icy bodies are active and evolving worlds. At the same time, major questions remain unresolved, including the chemical properties of icy materials, the geological histories of their surfaces, and the coupling between internal evolution and orbital dynamics. Current knowledge of the surfaces, interiors, and atmospheres of the principal icy bodies is built on spacecraft measurements, telescopic observations, laboratory experiments, and theoretical modeling. Recent contributions from Juno, James Webb Space Telescope (JWST), and stellar occultation studies have added valuable constraints on atmospheric composition, interior structure, and surface activity. Looking ahead, missions such as JUpiter ICy moons Explorer (JUICE), Europa Clipper, Dragonfly, and the Uranus Orbiter and Probe are expected to deliver substantial progress in the study of icy bodies. Their findings, combined with continued Earth- and space-based observations and laboratory studies, will be critical for assessing the potential habitability of these environments and for placing them within a broader framework of planetary system formation and evolution.
\end{abstract}

\noindent\textbf{Keywords}: satellites, geology, tectonics, atmosphere, interior structure, astrobiology, exploration, observation

\section{Introduction}
Ice typically refers to solid H$_{2}$O and is widely identified on Earth. In planetary science, ice refers to any solid volatile material, including not only water ice but also CO$_{2}$ ice, N$_{2}$ ice, hydrocarbon ice, and others. Water ice is the most common solid volatile material and has been confirmed to exist on various planetary bodies throughout the Solar System, even at the polar craters of Mercury. These various types of ice significantly influence the geological and atmospheric characteristics of solar system bodies. 

Only recently have scientists begun to understand the extensive distribution of ice throughout the solar system. \citet{Pilcher1972} was the first to detect H$_{2}$O on the surface of icy Galilean moons, Europa, Ganymede and Callisto, from ground-based observations. For the Saturn system, \citet{Johnson1975} found water ice on its moon Rhea first. However, the reflectance spectrum does not completely match that of pure water ice, indicating a presence of something other than pure water ice \citep{Kieffer1974}. Almost all solid bodies beyond the snow line (also called the ice line or frost line), which in the Solar System lies between the orbits of Mars and Jupiter and refers to the distance from a star beyond which temperatures are low enough for water vapor to condense into solid ice grains rather than remain as gas, are covered by ice on their surfaces and are collectively referred to as "icy bodies" (excluding the entirely rocky Jovian moon Io). (Figure 1). The densities of these bodies suggest that most of icy bodies have large amounts of water - although Europa with its relatively high density is exceptional (Figure 2), which suggests a large silicate fraction (Table 1).

Various geologic features exist on the surfaces of these icy bodies: narrow linear cracks on Europa and Enceladus, band-shaped arrays of grooves and ridges on Ganymede, Tethys and Miranda, and chaotically disrupted regions on Europa. Although impact craters are usually present on the surfaces, their abundance varies significantly among bodies, ranging from very few on Europa, which indicates a surface age on the order of tens of millions of years, to an almost globally saturated distribution on Callisto, whose surface age exceeds 4 billion years. The craters on icy bodies are generally shallower than those on rocky bodies, such as the Moon and Mercury, due to the higher mobility of ice. Additionally, on Enceladus, Triton, and possibly Europa, volatile materials erupt from the surface, known as cryovolcanoes which are similar to silicate volcanism but occur at much lower temperatures. Despite their cold environments far from the Sun, icy bodies exhibit a wide variety of geological activities and the remnants of such processes.

Observations for several icy bodies suggest the existence of a global liquid hydrosphere, or "subsurface ocean", beneath the ice. A potentially habitable environment may exist where various materials are dissolved in the geothermally heated water \citep{McMahon2021}. The presence and importance of water ice, along with other types of ices, within the solar system for surface tectonics, interior structure, atmosphere and evolution will be explored in this article. The insights presented are drawn from three decades of planetary exploration by spacecraft, supplemented by ground-based and space-based observations.

\begin{table}[t]
\centering
\caption{Fundamental Parameters for the icy bodies in the Solar System.}
\label{tab:table1}
\begin{tabular}{l l l l l}
\hline
System & Object & Mass & Mean radius & Mean density \\
 &  & ($10^{20}$ kg) [1] & (km) [1] & (kg m$^{-3}$) [1] \\\hline
Jupiter & Europa    & 480.0  & 1560.5 & 3010 \\
 & Ganymede  & 1481.9 & 2631.2 & 1940 \\
 & Callisto  & 1075.9 & 2410.3 & 1830 \\
Saturn  & Mimas     & 0.379  & 198.2  & 1150 \\
  & Enceladus & 1.08   & 252.1  & 1610 \\
  & Tethys    & 6.17   & 531.1  & 985  \\
  & Dione     & 11.0   & 561.4  & 1480 \\
  & Rhea      & 23.1   & 763.5  & 1240 \\
  & Titan     & 1345.5 & 2574.7 & 1880 \\
  & Hyperion  & 0.056  & 135.0  & 550  \\
  & Iapetus   & 18.1   & 734.4  & 1090 \\
  & Phoebe    & 0.083  & 106.5  & 1640 \\
Uranus  & Miranda   & 0.66   & 235.8  & 1201 \\
  & Ariel     & 12.3   & 578.9  & 1517 \\
  & Umbriel   & 12.8   & 584.7  & 1539 \\
  & Titania   & 34.5   & 788.9  & 1680 \\
  & Oberon    & 31.1   & 761.4  & 1682 \\
Neptune & Proteus   & 0.5    & 210.0 [2] & 1289 [2] \\
 & Triton    & 214.0  & 1353.4 & 2050 \\
 & Nereid    & 0.3    & 178.5 [3] & 1259 [3] \\
Dwarf planets & Pluto   & 130.3 & 1188.0 & 1854 \\
 & Charon  & 15.9  & 606.0  & 1700 \\
 & Haumea  & 40.1 [4] & 779.6 [4] & 2018 [4] \\
 & Makemake & 32.8 [5] & 715.0 [5] & 2140 [5] \\
 & Eris    & 163.8 [6, 7] & 1163.0 [8] & 2486 [6--8] \\
\hline
\end{tabular}

\vspace{2mm}
{\footnotesize
[1] Source: NASA Space Science Data Coordinated Archive https://nssdc.gsfc.nasa.gov. 
[2] \citet{Karkoschka2003}.
[3] \citet{Kiss2016}.  
[4] \citet{Dunham2019}.  
[5] \citet{Brown2013}.  
[6] \citet{Holler2021}.  
[7] \citet{Brown2023}. 
[8] \citet{Sicardy2011}. 
}\end{table}

\begin{figure}[htbp]
  \centering
  \includegraphics[width=1.0\linewidth]{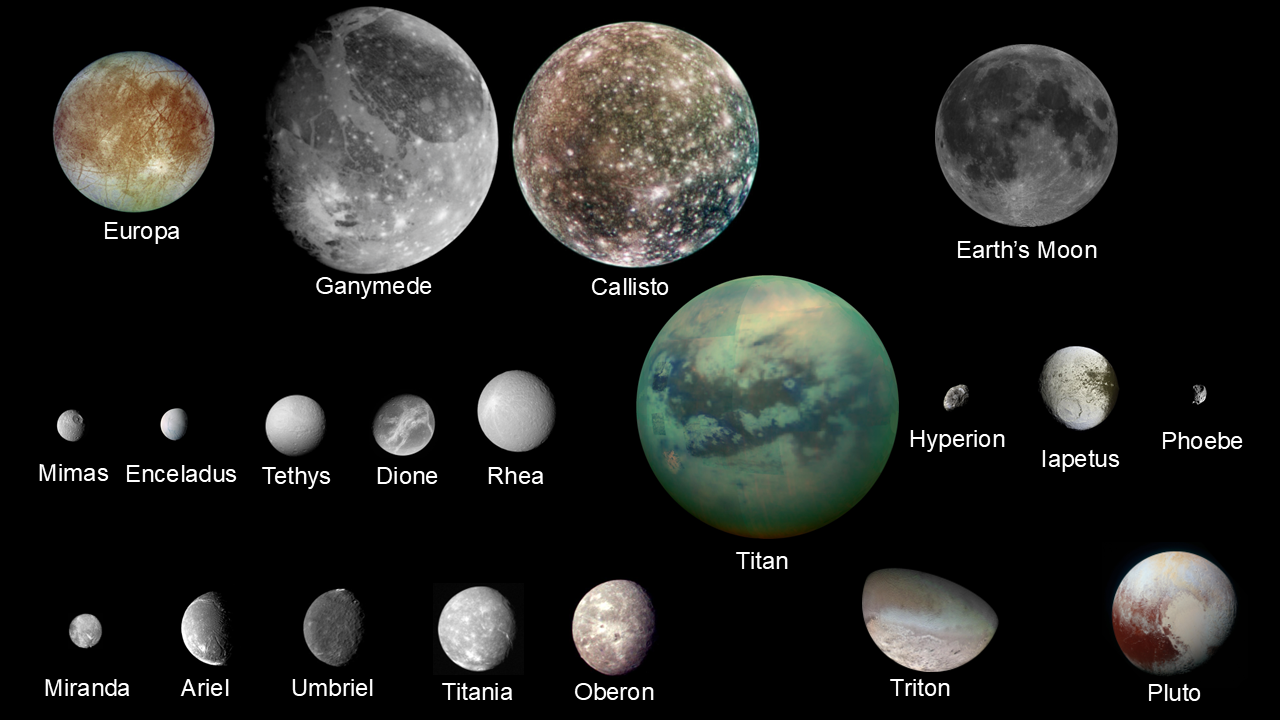}
  \caption{Major icy bodies in the Solar System, with the Earth's moon to scale. The images are shown at the proper relative size. From NASA's Planetary Photojournal (https://photojournal.jpl.nasa.gov/), Earth's Moon (PIA00302), Europa (PIA00502), Ganymede (PIA00716), Callisto (PIA03456), Mimas (PIA06258), Enceladus (PIA06254), Tethys (PIA07738), Dione (PIA08256), Rhea (PIA08256), Titan (PIA08256), Hyperion (PIA07740), Iapetus (PIA08384), Phoebe (PIA06064), Miranda (PIA18185), Ariel (PIA01534), Umbriel (PIA00040), Titania (PIA01979), Oberon (PIA00034), Triton (PIA00317), Pluto (PIA19952).}
  \label{fig:figure1}
\end{figure}

\begin{figure}[htbp]
  \centering
  \includegraphics[width=1.0\linewidth]{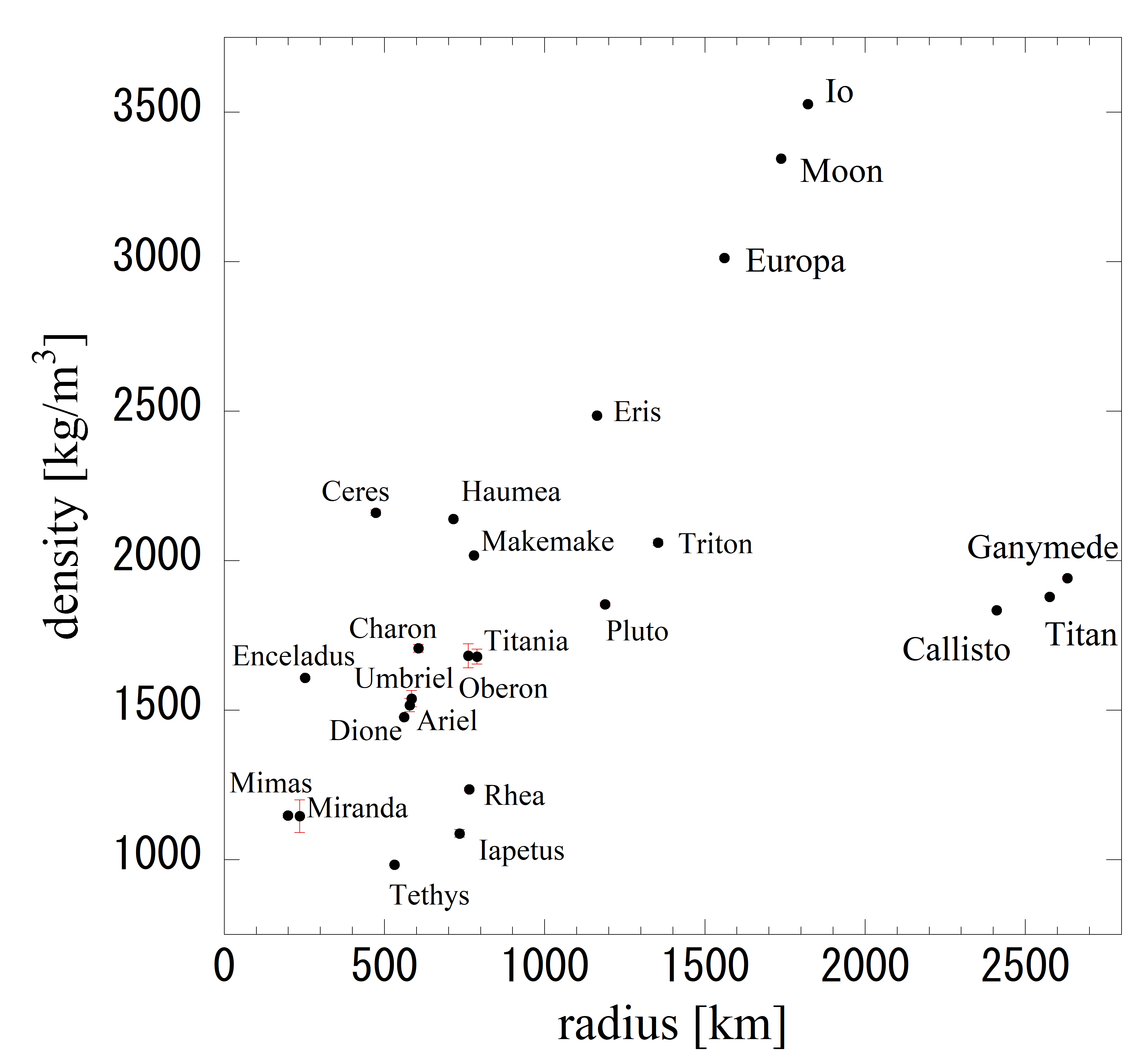}
  \caption{Radius-density relation of the icy bodies, Io and the terrestrial Moon.}
  \label{fig:figure2}
\end{figure}

%
\section{General description: Size, density, rotation, and interior structure}
The most basic physical characteristics of planetary objects are their size, mass, and bulk density derived from these quantities. Bulk density provides a first-order estimate of composition, since comparison with the densities of rock and water for icy bodies allows inference of the approximate rock-to-water ratio. However, density alone does not reveal whether the body's interior is differentiated into layers or how mass is distributed. For this reason, measurements of the gravity field and shape are critical. Second-degree gravity coefficients, in particular J$_{2}$ and C$_{22}$, allow an estimate of the moment of inertia  under the assumption of hydrostatic equilibrium (see e.g. \citet{Breuer2023} for more details). The moment of inertia reflects the degree of central mass concentration, with values smaller than 0.4 indicating a centrally condensed structure.

Tidal deformation provides an additional diagnostic. The tidal response is typically described by the tidal Love numbers, k$_{2}$, h$_{2}$, and l$_{2}$, first introduced by A. E. H. Love in his theory of elastic deformation of the Earth \citep{Love1892}. These parameters characterize different aspects of planetary response to tidal potential: k$_{2}$ measures potential perturbation in the gravitational field, h$_{2}$ represents vertical surface displacement, and l$_{2}$ describes horizontal displacement.
Tidal responses depend on the body's internal structure and composition; if it contains highly deformable materials, such as a global subsurface ocean, that respond strongly to tidal potential, the magnitude of the deformation can be substantially greater.
For Titan, k$_{2}$ has been derived from gravitational measurement by the Cassini spacecraft, leading to the prediction of a subsurface ocean \citep{Durante2019,Goossens2024}.

Many moons also exhibit synchronous rotation, in which orbital rotation is gravitationally locked to the central planet so that the same hemisphere always faces the planet. Such a body has leading and trailing hemispheres relative to the orbital direction. Synchronous rotation leads to triaxial deformation under the combined effect of self-rotation and tidal forces, and precise measurements of a moon's shape can provide further constraints on its interior. If the orbital periods of two moons can be expressed as a ratio of integers, they are in orbital resonance. Examples include the 1:2:4 resonance of Ganymede, Europa, and Io, and the 1:2 resonance of Dione and Enceladus. Orbital eccentricity of moons in resonance can be strongly enhanced, and the resulting frictional heat from periodic deformation can become a significant internal energy source. Orbital eccentricity also induces oscillations in a moon's spin rate about its equilibrium mean value, producing a forced libration primarily in longitude. Measurements of libration amplitudes can provide evidence for subsurface oceans, as suggested for Enceladus \citep{Thomas2016}.

Electromagnetic interactions between a planet and its moon can provide crucial information about the presence of a global salty ocean within the moon. If the planet's magnetospheric equator is tilted relative to the moon's orbital plane, the moon passes through both the northern and southern hemispheres of the planet's magnetosphere during its orbit. These periodic shifts in the magnetic field generate eddy currents in any conductive layer within the moon, such as a subsurface salty ocean, which in turn induces a secondary magnetic field. At Europa, Ganymede, and Callisto, Galileo spacecraft observed fluctuations in the surrounding magnetic environment, which strongly suggest the presence of subsurface salty oceans \citep{Kivelson2007}. However, if the planet's magnetospheric equator is aligned with the moon's orbital plane, as is the case in the Saturnian system, the magnetic field at the moon remains constant over time, preventing any significant inductive response.

Impact craters on surfaces provide crucial information for understanding the geological history of these bodies. Generally, crater size-frequency distributions (SFDs) on bodies in the outer Solar System differ significantly from those observed in the inner Solar System such as on the Moon (Strom et al., 2015). In the outer Solar System. the main sources of impactors are considered to be ecliptic comets and long-period comets \citep{Levison2000,Zahnle2003}, which contrasts with the small asteroids from the main asteroid belt that primarily strike terrestrial bodies in the inner Solar System \citep{Strom2015}. As a result, impact crater production functions derived from the Moon and Mars are inconsistent with those on the outer Solar System. It should be noted that surface ages estimated from crater chronologies for outer Solar System bodies include large uncertainties in impact rates and crater formation mechanisms.

\begin{figure}[htbp]
  \centering
  \includegraphics[width=1.0\linewidth]{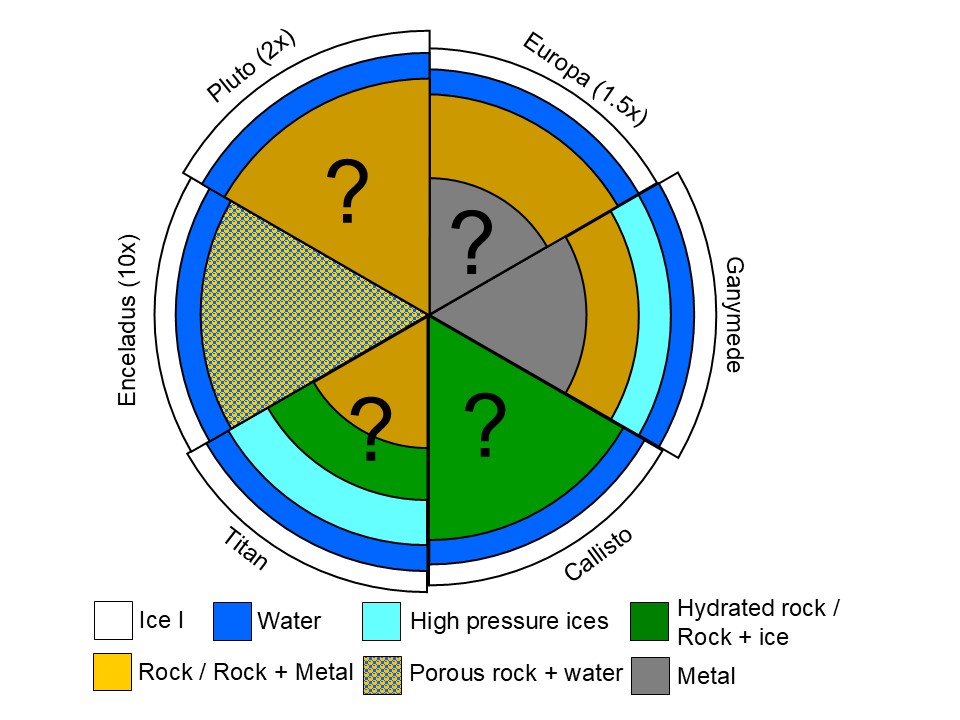}
  \caption{Inferred internal structures for selected icy bodies. Note that Ganymede, Callisto, and Titan are shown at their correct relative sizes, while Europa, Pluto and Enceladus are plotted 1.5, 2, and 10 times their actual sizes, respectively. Also see Figure 1 and Table 1 for a size comparison of each object.}
  \label{fig:figure3}
\end{figure}

%
\section{Characteristics of the icy bodies}
This chapter provides an overview of current knowledge of icy bodies derived from observations, in-situ explorations, and theoretical modeling. For each body, recent results and progress are also summarized.
\subsection{Europa}
Europa, the smallest of the Galilean satellites with a radius of 1,561 km, orbits Jupiter in 3.55 days at an average distance of 671,000 km (9.4 times Jupiter's radius). 
With a bulk density of 3010 kg/m$^{3}$, it is estimated that Europa contains at least several percent water by mass, an amount exceeding the total volume of Earth's surface water. 
Based on the bulk density and measurements of Europa's gravity field, Europa's interior is likely to have a three-layer structure: a metallic core at its center, a rocky mantle surrounding the core, and an overlying hydrosphere with a thickness of 200 km or less. 
The core is estimated to be between 40\% and 50\% of Europa's radius (Figure 3). 
Magnetic field data strongly suggests that Europa possesses an induced magnetic field, likely generated by a subsurface layer of conductive material, which may be a global ocean of salty liquid water \citep{Kivelson2007}.

Europa's surface has a high albedo, and its global spectral signature closely matches that of pure water ice. 
Albedo and color variations, both on hemispheric and local scales, are striking features of its surface. 
The trailing hemisphere is notably darker than the leading hemisphere, likely to be influenced by the bombardment of charged particles from Jupiter's magnetosphere, which can alter the surface composition and structure of the ice. 
SO$_{2}$ has been detected mainly on the trailing hemisphere, where charged particle flux is strongest because these particles nearly co-rotate with Jupiter, consistently overtaking Europa in its orbital path \citep{Lane1981,McEwen1986}. 
This sulfur is thought to originate from Io's volcanic eruptions. 
On a smaller scale, reddish-brown material coats fractures and other geologically young features. Spectroscopic studies suggest that these reddish deposits may be rich in salts such as magnesium sulfate or carbonate, likely left behind by evaporating briny water that has seeped from beneath the surface \citep{McCord1999}. 
Another possible cause of brownish discoloration is sodium chloride. Laboratory experiments show that irradiated sodium chloride, exposed to conditions equivalent to 10 to 100 years on Europa's surface, can explain the visible spectrum detected by the Solid-State Imaging (SSI) system \citep{Hand2015}.

A report of a possible water plume from the southern hemisphere detected from ultraviolet observations by the Hubble Space Telescope (HST) \citep{Roth2014a} suggests that material from Europa's subsurface ocean could be ejected into space, potentially providing direct access to this material through the ice shell. 
Europa's atmosphere is also replenished by materials stripped from its icy surface due to the effects of Jovian plasma radiation, thermal desorption by solar radiation, and meteoritic impacts. 
This process is estimated to populate water, oxygen, and hydrogen molecules at a rate of about 10$^{26}$-10$^{27}$ molecules per second \citep{Vorburger2018}. 
The estimated abundances of water vapor from the plumes could exceed those generated by exogenic effects by factors of 10--100. 
However, excepting one tentative detection of water out of 17 observations from infrared spectroscopy using the Keck telescope \citep{Paganini2020}, subsequent observations have not confirmed this plume activity, despite several attempts using HST \citep{Roth2014b}, 
Subaru Telescope \citep{Kimura2024} and James Webb Space Telescope \citep{Villanueva2023a}. These results support the hypothesis that potential plumes on Europa may be confined to specific areas, occur infrequently, exhibit low intensity, or contain minimal volatile components. 
Even if present, the details regarding their temporal and spatial variations remain unclear. 
In 2030's, Europa Clipper will search the plume during multiple flybys to Europa \citep{Pappalardo2024}. 
If the plume really exists, in-situ measurements expect to clarify potential plume sources and understand their extent, distribution, and variability.

Materials sputtered from its icy surface create an exceedingly tenuous atmosphere, with a surface pressure of just 0.1 $\mu$Pa; roughly 10$^{-12}$ times that of Earth's atmosphere \citep{Hall1995}. The atmosphere consists mainly of molecular oxygen, which is thought to result from the dissociation of surface water ice by solar ultraviolet radiation and energetic particles from Jupiter's magnetosphere. Recent in-situ measurements by Ju-no detected pickup ions H$_{2}^{+}$ and O$_{2}^{+}$ near Europa, providing the first direct sampling of key atmospheric constituents and constraining surface oxygen production to 12 $\pm$ 6 kgs$^{-1}$ (2.2 $\pm$ 1.2$\times$10$^{26}$ s$^{-1}$) which is markedly lower than some earlier model estimates, with implications for redox delivery to the subsurface ocean and overall habitability assessments \citep{Szalay2024a}. 
At the same time, other Juno observations have shown that Europa measurably modifies the composition and dynamics of Jupiter's plasma sheet, demonstrating that it remains a significant source of mass and energy to the magneto-sphere \citep{Szalay2024b}. 
These results together suggest that even a modest neutral source from Europa can exert a notable influence on the Jovian magnetospheric environment.

Europa's surface has a notable scarcity of large impact craters. 
The overall trend of Europa's crater size-frequency distributions (SFDs) is significantly different from those observed in the inner solar system (e.g., the Moon) \citep{Strom2015}. 
The impacting population on Europa (and other Galilean moons) consists mainly of Jupiter-family comets and long-period comets \citep{Levison2000,Zahnle2003}, which contrasts with the small asteroids from the main asteroid belt that strike terrestrial bodies in the inner solar system \citep{Strom2015}. 
Consequently, impact crater production functions derived from the Moon and Mars are inconsistent with those on Europa. 
The average surface age of Europa is estimated to be between 20 and 200 million years, with uncertainties stemming from the comet impact rate and cratering mechanisms \citep{Bierhaus2013}. 
This places Europa's surface as relatively young on solar system timescales. 
Despite the lack of large craters, the shapes and morphologies of existing craters provide insights into Europa's internal structure and geological activity. 
The relationship between crater depth and diameter on Europa shows that craters larger than approximately 3 km in diameter are shallower than those on the Moon, and craters larger than 8 km exhibit decreasing depth with increasing diameter. 
This reverse trend is likely due to the excavation of material from a weak, highly mobile layer of warm ice, which flattens the larger craters \citep{Schenk2009}.

Europa's surface showcases a range of geological features, indicating significant tectonic activity. Lineae, the most dominant feature, are elongated surface formations that can form ridges, often appearing as "double ridges" with a central trough, which range from $\sim$200 m to over 4 km wide and can extend for more than 1,000 km (Figure 4a). One model suggests these ridges form when tidal forces crack the ice, allowing subsurface water to rise and freeze \citep{Greenberg1998,Tufts2000}. 
Another theory proposes that ridges form through horizontal motion along fractures, generating shear heating that raises warm ice \citep{Gaidos2000,Nimmo2002}. 
In particular, arcuate ridge segments linked at sharp cusps are also known as flexus or cycloids. 
These curved tensile fractures are thought to result from the rotation of Europa's diurnal stress field. 
In addition, the gradual solidification of the subsurface ocean and the accompanying increase in volume generate sufficient tensile stress on the surface, which could contribute to the formation of linear-shaped features \citep{Kimura2007,Manga2007}. 
Chaos Terrain consists of regions of disrupted surface where polygonal blocks of pre-existing ice plains are surrounded by mound-shaped material. 
Conamara Chaos is a prime example, showing tilted and rotated terrain blocks (Figure 4b). 
Two models explain the formation of Chaos Terrain: the Melt-Through model suggests localized ice melting from Europa's mantle \citep{Thomson2001,OBrien2002}, while the Brine Mobilization model proposes that antifreeze materials lower ice's melting point, creating the observed topography \citep{Schmidt2011}. 
Dark spots with varied shapes (circular, elliptical, or irregular) are called Maculae, which are more likely regions of chaos terrain that have sunk below the surrounding surface. 
Thera Macula, for example, may harbor a subsurface liquid lake, as suggested by the Melt-Lens model \citep{Schmidt2011} (Figure 4c). 
Large Ringed Features are circular formations over 100 km wide, thought to be ancient impact sites that have penetrated Europa's subsurface layers (Figure 4d). 
Europa's geological features are influenced by global stress mechanisms and subsurface processes. 
Data from the Voyager and Galileo missions suggest subduction-like activity, where new surface material is recycled into the ice shell at compressional bands, akin to Earth's tectonic behavior. 
Although the thickness of the ice shell has been estimated to range from a few km to about 40 km \citep{Schubert2009,kimura2024europa}, its exact value and physical/chemical properties remain uncertain. 
Future missions are crucial to further understanding these dynamics and Europa's geological evolution \citep[e.g.][]{Pappalardo2024}.
\begin{figure}[htbp]
  \centering
  \includegraphics[width=1.0\linewidth]{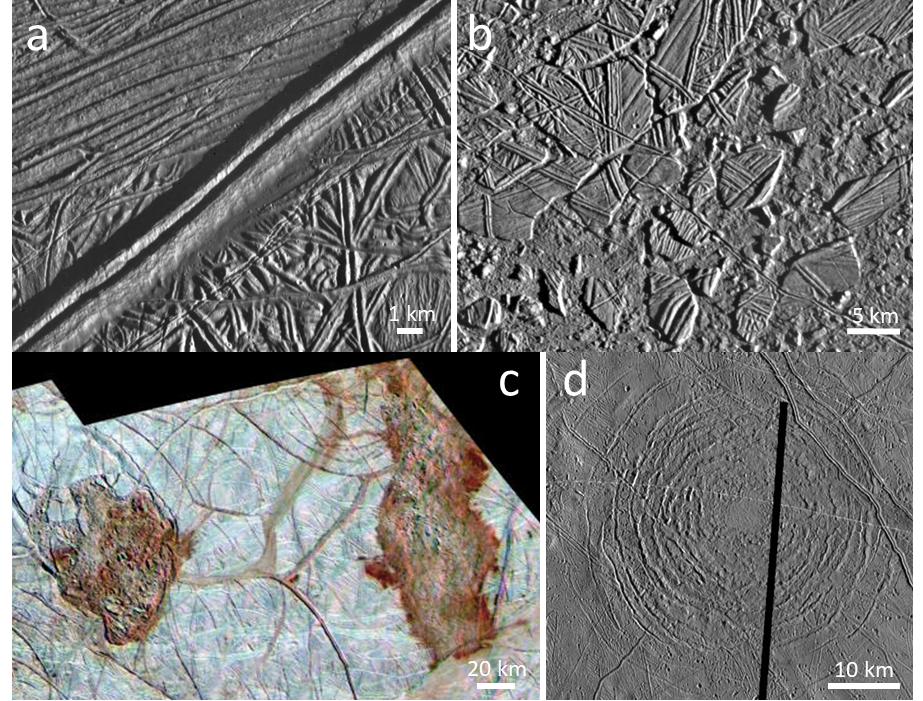}
  \caption{Four types of surface features on Europa. (a) double ridge feature (PIA00589), (b) Conamara chaos (PIA00591), (c) Thera (left) and Thrace (right) macula showing dark, reddish disrupted regions (PIA02099), and (d) Large ringed feature, Tyre (PIA01661).}
  \label{fig:figure4}
\end{figure}
%
\subsection{Ganymede}

With a radius of 2631.2$\pm$1.7 km \citep{Archinal2018}, Ganymede is the largest satellite in the Solar System. Gravity field data obtained by the Galileo spacecraft, and more recently by Juno, along with spectral surface data, suggest that Ganymede's interior is composed of water ice and rock-metal components in a mass ratio of approximately 40\% H$_{2}$O to 60\% rock and metal. 
Ganymede is highly differentiated \citep{Schubert2004,GomezCasajus2022,Hussmann2025}. 
Notably, Ganymede possesses an intrinsic magnetic field, likely generated by dynamo action in a liquid Fe-FeS core \citep{Schubert1996,Kivelson2002}. 
Models that account for Ganymede's mean density of 1942.0$\pm$4.8 kg/m$^{3}$ and moment of inertia of 0.3105$\pm$0.0028 suggest an interior structure consisting of an iron-rich core, a surrounding silicate rock mantle, and an overlying ice shell. 
A possible core radius depends on the relative proportions of major light elements and remains poorly constrained, but the moment of inertia factor restricts the core radius to a range between 600 km for a pure iron core and 1150 km for a FeS core (Figure 3). 
The thickness of the water layer is estimated to range from 1000 to 1100 km above the silicate mantle \citep{Anderson1996,Sohl2002,Schubert2004}. 
An observed inductive response is consistent with the presence of a buried conducting shell, likely a subsurface liquid water ocean, though other interpretations cannot be ruled out. 
The large thickness of Ganymede's water layer makes its structure significantly different from that of Europa. 
Due to the high pressure at the bottom of the water layer, high-pressure ice phases must exist below a depth of approximately 150 km. 
Thus, the water layer is subdivided into Ice Ih, Ice III, Ice V, and Ice VI as pressure increases. High-pressure ices are solid forms of water that occur only under pressures much greater than those at Earth's surface, and they have denser crystal structures than ordinary ice (Ih). For example, Ice Ih is the hexagonal ice familiar on Earth, stable at low pressures (leas than $\sim$0.2 GPa) with a density of ~920 kg/m3, whereas Ice III and Ice V are stable at intermediate pressures ($\sim$0.2-0.6 GPa) and are denser ($\sim$1140-1240 kg/m$^{3}$), and Ice VI forms at still higher pressures ($\sim$0.6-2.2 GPa) with a density near 1310 kg/m3. If a liquid water ocean exists, as indicated by magnetic data, it would likely be located between the Ice Ih and the high-pressure ice layers. 

The surface of Ganymede is divided into two main terrain types: relatively old, dark (heavily cratered) terrain, and relatively younger, cross-cutting lanes of bright (typically grooved) terrain (Figure 5).
Dark terrain covers about one-third of Ganymede's surface and is estimated to be older than 4 billion years, based on crater size and density relationships. The smallest craters are simple bowl-shaped, but as crater size increases, the morphology transitions into more complex forms. In larger craters, the morphology evolves into central pits, central domes, penepalimpsests, palimpsests, and multi-ring structures with increasing diameter. Particularly vast multi-ring structures, ranging from tens to hundreds of kilometers, are thought to represent fault-bounded troughs, termed furrow (fossae) systems, formed in response to large impact events into a thin lithosphere in the early stages of Ganymede's history \citep[e.g.][]{Hirata2020}. The dark terrain is typically modified by such furrows \citep{Pappalardo2004,Prockter2010}.

The bright terrain covers the remaining two-thirds of the surface and separates tracts of dark terrain in wedge-shaped forms. It is ubiquitously grooved, indicating episodes of widespread extensional tectonism. Although the bright terrain is still fairly cratered, it is clearly younger than the dark terrain, as evidenced by the truncation relationships with dark terrain and the lower crater density. However, estimating the precise age of the bright terrain is difficult due to the low resolution of available images and uncertainties in impact flux. High-resolution images reveal that the bright terrain consists of belts of subparallel grooves, which likely formed due to tilt-block normal faulting of the surface's brittle icy layer. The large-scale topography may be caused by the necking of the brittle layer over a ductile substrate \citep{Pappalardo1998,Giese1998,Prockter1998,Head2002}.
Possible explanations for the origin of the bright terrain include global expansion driven by internal differentiation, subsequent heating, and phase transitions in solid ice. 
Additionally, relatively smooth areas are also present, which may have formed through cryovolcanic processes \citep{Prockter2010}. 
Nevertheless, the current data lack the necessary resolution and coverage to definitively support this cryovolcanism hypothesis.
\begin{figure}[htbp]
  \centering
  \includegraphics[width=1.0\linewidth]{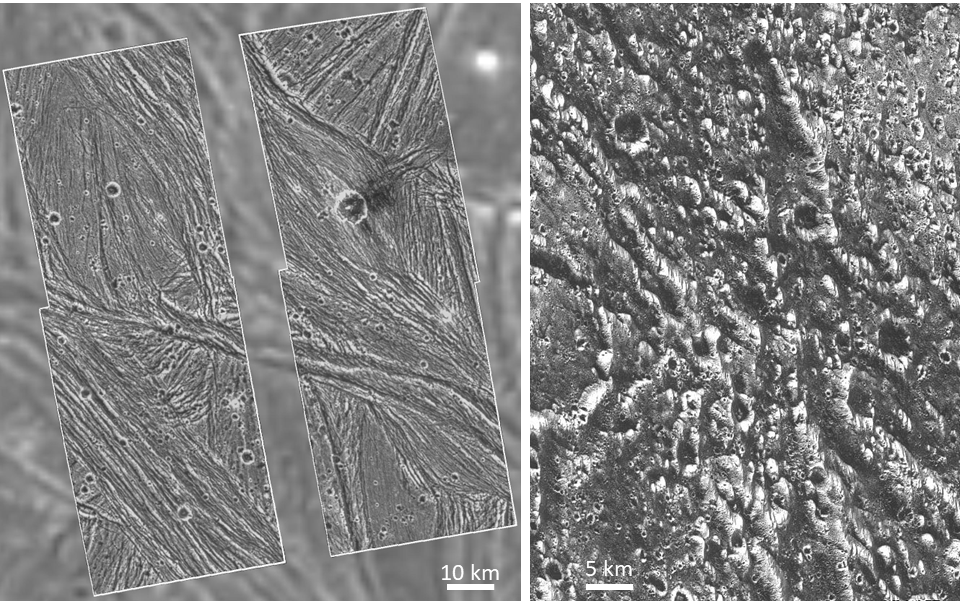}
  \caption{(Left) Grooved terrain on Ganymede. A mosaic of Galileo spacecraft high-resolution images of the Uruk Sulcus region is shown within the context of an image of the re-gion taken by Voyager 2 (PIA00281). (Right) Cratered terrain on Ganymede. Ancient impact craters shown in the image taken by Galileo spacecraft (PIA00279).}
  \label{fig:figure5}
\end{figure}
%
\subsection{Callisto}

Whereas Ganymede and Europa show regions that have been modified extensively by endogenic geological processes, the surface of Callisto is dominated by craters of various sizes which indicate a surface age exceeds 4 billion years \citep{Zahnle2003,Moore2004}. 
The impact craters exhibit a wide range of morphologies, from small, with diameters less than 5 km, simple bowl-shaped craters to over several hundreds kms in diameter, massive multi-ring structures. Crater 5-40 km across typically has a central peak and ranging from 25 to 100 km in diameter often have central pits instead of peaks. 
Over 100 km in diameter, exhibit bright, shallow, dome-shaped structures. The largest multi-ring basin, Valhalla, with a bright central region 600 km in diameter and rings ex-tending up to 1,800 km from the center likely formed as a result of concentric fracturing of the lithosphere, which lies atop a layer of soft or liquid material, possibly an ocean, following an impact (Figure 6).

Surface features on Callisto have been almost entirely modified due to weather-ing processes and erosion. Palimpsests are unique impact structures on Ganymede and Callisto that are devoid of typical crater morphologies such as crater rims \citep{Jones2003,Pappalardo2004}. 
Fluid-rich continuous ejecta was formed due to impact into an icy surface and then the ejecta deposition and the gravitational relaxation removed the topography of the crater cavity \citep{Jones2003}. In addition, Callisto's surface is notably devoid of small craters less than 1 km in diameter. Instead, the surface is covered with small pits and protrusions. This is thought to be the result of ice sublimation at temperatures of around 165 K at the subsolar point, leaving behind debris from other materials on the crater rims. Callisto's surface is one of the darkest among the icy moons, with an albedo of approximately 0.2. This low albedo leads to increased heating due to solar insolation, which in turn creates a feedback loop that sublimates the brighter ice and in-creases the amount of dark non-icy remnants, making the surface even darker.

A relatively large moment of inertia factor (0.3549$\pm$0.004221) suggests that Callisto's interior is only partially differentiated, meaning it is composed of ice and rock, with the proportion of rock increasing with depth (Figure 3). 
However, the precise value of the moment of inertia factor itself remains under debate, partly because non-hydrostatic contribu-tions to Callisto's gravity field may not be negligible \citep{Gao2013}. 
As a result, the degree of differentiation is uncertain, and competing interior models remain under discussion. 
These range from a largely undifferentiated interior with a thick ice-rock mixed mantle, to configurations with a small metallic core, an ice-rock mixture at intermediate depths, and an overlying water-ice layer that may include a liquid ocean. Although an inductive magnetic signature has been detected, suggesting the presence of a subsurface liquid water ocean, its depth and thickness remain unconstrained, and there are no clear signs of tectonics or resurfacing. 
Callisto lacks an intrinsic dipole magnetic field, consistent with the absence of a fully differentiated metallic core.
\begin{figure}[htbp]
  \centering
  \includegraphics[width=1.0\linewidth]{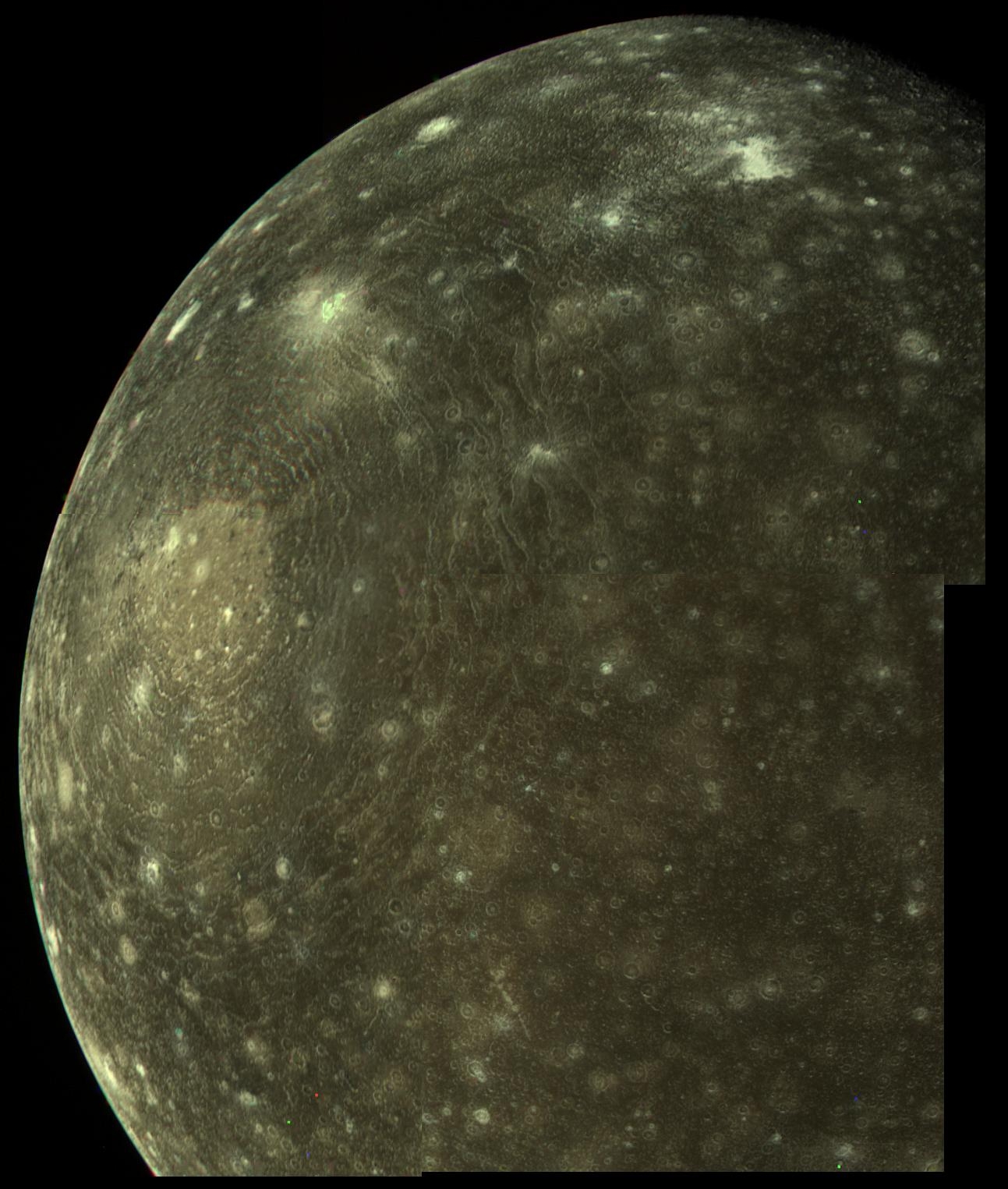}
  \caption{Mosaic image of Callisto obtained by Voyager 1 spacecraft shows surface detail as small as 10 kilometers across. The prominent old impact feature (palimpsest) Valhalla has a central bright spot about 600 kilometers across (PIA00080).}
  \label{fig:figure6}
\end{figure}
%
\subsection{Mimas and Enceladus}
Mimas has 198 km in radius, which is the smallest moon known to be roughly rounded shape due to its own gravity. Bulk density of 1,150 kg/m$^{3}$ indicates that it is composed mostly of water ice with only a small amount of rock. Although its surface is heavily cratered and shows little signs of recent geological activity, recent dynamical analyses indicate that Mimas likely harbors a present-day global ocean at $\sim$20-30 km depth, with an inferred age less than 25 Myr \citep{Lainey2024}. 
The analyses suggest that Mimas, previously regarded as geologically inert, could instead be a nascent ocean world, broadening target lists for comparative ocean-world studies. A notable feature on its surface is Herschel crater, 139 km across which is roughly one-third of Mimas's mean diameter, and one of the largest craters in the Solar System relative to the size of its parent body. 

Enceladus is the sixth largest of Saturn's moons (with a radius of 252 km), orbiting Saturn at a distance of about 238,000 km (3.95 Saturn radii) every 1.4 days, with the same spin period. Average density of 1610 kg/m$^{3}$ indicates that approximately 40\% of Enceladus is composed of H$_{2}$O. The moment of inertia factor is 0.3305, significantly lower than the 0.4 value expected for a homogeneous interior, suggests that Enceladus' interior is differentiated into rock and H$_{2}$O . If the rock has a density of 2,450 kg/m$^{3}$, the rocky core is estimated to be about 190 km radius (Figure 3) \citep{McKinnon2015,Hemingway2018}. 
Enceladus' surface is covered by a solid H$_{2}$O and shows strong het-erogeneity between the northern and southern hemispheres. The northern hemisphere is heavily cratered and extremely old (approximately 4 billion years), while the southern hemisphere has far fewer impact craters and is extremely younger (between 1 million and several hundred million years) (Patterson et al., 2018). In the southern hemisphere, many linear ridges and grooves, similar to those seen on Jupiter's moon Europa, are present, with the largest reaching 200 km in length, 10 km in width, and 1 km in height (or depth). This indicates that the ice shell has experienced stress, likely caused by convection within the ice shell or periodic tidal deformation. Near the south pole, several linear features known as "tiger stripes" are almost devoid of impact craters \citep{Spencer2018} (Figure 7). 
The maximum temperature in this region reaches about 160 K, significantly higher than the average surface temperature of 75 K. The heat flux from this area is around 200 mW/m$^{2}$, more than 10 times higher than the heat expected from radioactive decay of rocks inside Enceladus \citep{Spencer2018}.

The Cassini spacecraft discovered that an active plume erupts from the tiger stripes in the south polar region. There are more than 100 vents along these features, ejecting approximately 250 kg of material per second, primarily solid and gaseous H$_{2}$O \citep{Postberg2018}. 
Some materials are ejected at speeds of up to 3 km/s, and those exceeding Enceladus' escape velocity contribute to Saturn's rings. JWST imaging captured an exceptionally extended water-vapor plume from Enceladus, orders of magnitude larger than the moon itself, and traced material feeding the E-ring torus. 
The observations indicate a plume mass flux consistent with Cassini-derived estimates and, importantly, provide independent confirmation while placing those earlier inferences into a broader spatial context of ocean-sourced material exchange \citep{Villanueva2023b}. 
The strength of the eruptions fluctuates, increasing near Enceladus' apogee and decreasing near perigee, likely due to changes in the stress on the linear features, which become tensile near apogee, widening the cracks, and compressive near perigee \citep{Spencer2018}. 
Ion and Neutral Mass Spectrometer (INMS) onboard Cassini spacecraft revealed that the gas components of the plume include H$_{2}$O, hydrogen, carbon dioxide, and carbon-bearing molecules such as methane, ammonia, propane, formaldehyde and phosphates \citep{Postberg2018,Postberg2023}. Solid components include H$_{2}$O ice, sodium salts, complex organic molecules containing aromatic hydrocarbons with molecular weights exceeding 200, and nanometer-sized silica particles \citep{Postberg2018}. The presence of sodium salts suggests that seawater interacts with rocks in Enceladus' rocky core, and the abundance of silica particles and hydrogen indicates that hydrothermal activity occurs on the seafloor, with temperatures exceeding 100 degrees Celsius \citep{Postberg2018}.

Enceladus' orbit is slightly eccentric, leading to slight oscillation (forced libration) in the orientation of its surface (ice shell) relative to its core. 
Such libration, with a magnitude of about 0.1 degrees, suggests that the surface ice shell is not rigidly attached to the rocky core, implying the existence of a global subsurface ocean beneath the ice shell. Based on these observations, along with plume composition data and geodetic data, it is believed that Enceladus has a subsurface ocean approximately 30 km thick beneath an ice shell about 25 km thick. 
Additionally, the Antarctic region ap-pears to be slightly de-pressed, suggesting that the ice shell there is thinner, around 10 km, and the subsurface ocean is correspondingly thicker. For an emergence of life, the coexistence of a "solvent", "energy", and "organic matter" is required \citep{McMahon2021}. 
Enceladus, with its subsurface ocean, sufficient energy for hydrothermal activity, and the detection of various organic compounds, appears to meet these conditions. Future studies will need to further investigate each condition and quantitatively analyze the potential chemical reactions and the amounts of products.
\begin{figure}[htbp]
  \centering
  \includegraphics[width=1.0\linewidth]{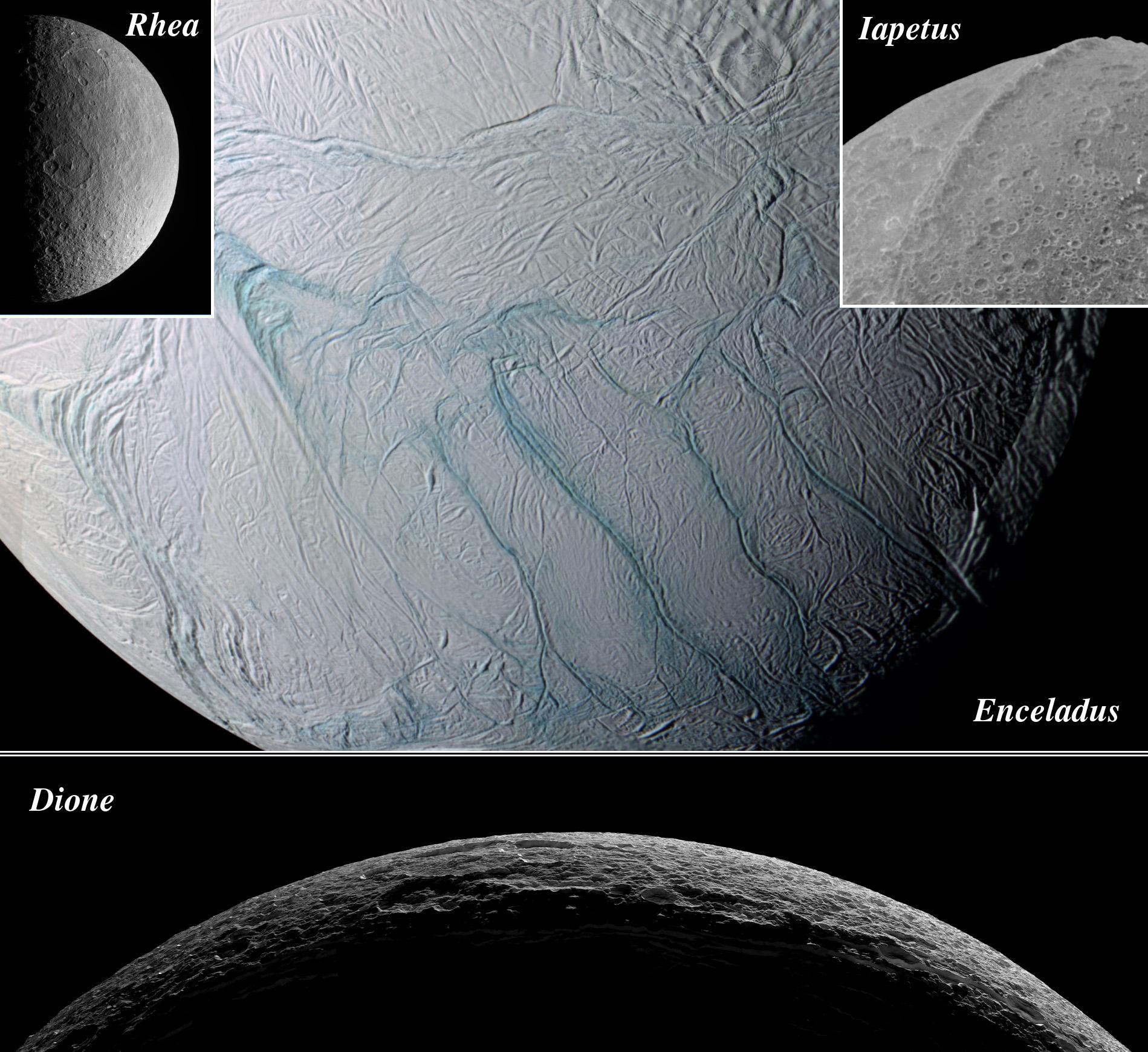}
  \caption{Four major icy moons of Saturn that the Cassini spacecraft visited during 2005 (PIA07767).}
  \label{fig:figure7}
\end{figure}
%
\subsection{Tethys, Dione and Rhea}

Tethys has a radius of 531 km and a bulk density of 985 kg/m$^{3}$, indicating that it is com-posed almost entirely of water ice. 
Because the Cassini spacecraft has not performed a close flyby of Tethys, it is not possible to estimate its moment of inertia based on the gravitational field measurement. Its surface is among the brightest in the Solar System, with a visual albedo of 1.23. 
This exceptionally high albedo is likely due to the deposition of water ice particles from Saturn's E-ring, which are generated by Enceladus' water plumes. 
Supporting this hypothesis is the observation that the reflectivity of Tethys' leading hemisphere is 10\% higher than that of its trailing hemisphere \citep{Filacchione2007}. 
While no materials other than water ice have been definitively identified on Tethys, possible constituents include ammonia, carbon dioxide, and organic compounds. The trailing hemisphere becomes darker and redder as it approaches the anti-apex, exhibiting the same spectral properties observed on the surfaces of Iapetus and Hyperion \citep{Jaumann2009}. 
The surface is mostly dominated by impact craters with an estimated age over 4 Gyrs. 
Largest impact basin, Odysseus, with 450\,km diameter has quite flat floor which is most likely due to viscous relaxation over geologic time. Estimated age for the youngest unit within the Odysseus is from 3.76 to 1.06 Gyrs \citep{Dones2009}. 
Also, a number of tectonic features such as chasmata and troughs intersect the cratered surface. Largest valley called Ithaca Chasma extends over 2,000 km in length, 100 km wide and 3 km deep, and is older than Odysseus \citep{Dones2009}. 
It is thought to be formed due to global expansion event in the past.

Dione has a mean radius of 561 km and a bulk density of 1,480 kg/m$^{3}$. An estimation of the gravity field of Dione based on the Cassini data acquired during five close flybys of the moon indicates a significant departure from the expectation for a body that has relaxed to hydrostatic equilibrium, which means that the moment of inertia cannot be inferred directly from the gravitational field \citep{Zannoni2020}. 
The surface is entirely covered with impact craters (Figure 7), with the trailing hemisphere having a higher crater density than the leading hemisphere. These craters are crossed by ice cliffs formed by tectonic fractures, which appear as bright streaks. Cassini spacecraft detected a very thin layer (exosphere) of molecular oxygen ions around Dione \citep{Tokar2012}. 
It seems that highly charged particles from the Saturn's strong radiation belts could split the water in the ice into hydrogen and oxygen.

Rhea is the second-largest moon of Saturn, with a radius of 764 km and a bulk density of 1,240 kg/m$^{3}$. This low density suggests that it is composed of roughly three-quarters of water ice, with the remainder being rock. Based on gravity measurements by the Cassini spacecraft, Dione's interior is likely in hydrostatic equilibrium, and its moment of inertia factor has been estimated to be approximately 0.37 \citep{Iess2007}. 
However, another analysis argued that Rhea may not be in hydrostatic equilibrium and that its moment of inertia cannot be determined solely based on gravity field data \citep{McKenzie2008}. 
Rhea's surface shows similarities to that of Dione, including heavily cratered terrain with distinct differences between the leading and trailing hemi-spheres (Figure 7). The leading hemisphere is bright and heavily cratered, while the trail-ing side exhibits bright tectonic features (such as graben and troughs) on a dark back-ground, which is the opposite trend observed on Dione's surface. Cassini data suggests tectonic activity, including ice cliffs that form bright lines. Evidence of internal activity, such as fault systems and large, relaxed craters (over 100 km in diameter), indicates past cryovolcanic processes and resurfacing. The dark regions are likely composed of tholins, which are formed through reactions involving carbon, nitrogen, and hydrogen \citep{Cruikshank2005}. 
The tenuous exosphere is composed of oxygen and carbon dioxide in a ratio of approximately 5 to 2. The primary source of oxygen is the radiolysis of water ice on the surface, caused by irradiation from Saturn's magnetosphere. The source of car-bon dioxide is less well understood; it may be linked to the oxidation of organics present in the ice or to outgassing from the moon's interior \citep{Teolis2010}.
%
\subsection{Titan}
Titan, another astrobiological target in the Saturn system, is quite different from Enceladus and Europa and is unique among the moons of the Solar System. Its most no-table feature is that it is the only moon in the Solar System with a thick atmosphere, with a surface pressure of about 1.5 bar. Titan orbits Saturn with a period of 15 days and 22 hours in synchronous rotation. With a radius of 2,575 km, Titan is the second largest moon in the Solar System after Jupiter's moon Ganymede, larger than the planet Mercury, which has a radius of 2,439 km \citep{Breuer2023}. 
Titan's surface is covered in ice, and its bulk density of 1,881 kg/m$^{3}$ suggests that Titan is composed of 40-60\% rock, with the rest being mainly water ice. Assumed that Titan's interior consists of two layers, an outermost pure water-ice shell and an inner constant-density core, then the bulk density and the moment of inertia factor of 0.348 determine the radius of the core to be 2,110 km~$\pm$~km and the density of the core to be 2,528\,$\pm$\,175 kg/m$^{3}$ \citep{Goossens2024}. 
Based on the gravitational measurements by the Cassini, the tidal Love number of 0.375\,$\pm$\,0.06 has been derived, which indicates the global ocean with a thickness of at least 50 km and a density of about 1020 kg/m$^{3}$ exists below the ice shell with a thickness of also at least 50 km thickness (Figure 3) \citep{Goossens2024}.

Titan's atmosphere, which extends to altitudes over 1,000 km, is primarily com-posed of N$_{2}$, and its surface pressure reaches 1.5 atm. In addition to N$_{2}$, Titan's atmosphere contains about 1.5 to 5\% CH$_{4}$, which undergoes photochemical reactions to produce hydrocarbons such as C$_{2}$H$_{6}$ and C$_{2}$H$_{2}$, as well as cyanide compounds like HCN and CH$_{3}$CN through reactions with N$_{2}$. Furthermore, in the ionosphere at altitudes around 1,000 km, polymerization reactions create organic molecules with molecular weights of several thousand. As these polymeric organic materials descend through the atmosphere, they form aerosol particles with diameters of several tens to several hundred nanometers. At an altitude of about 500 km, these particles aggregate into a dense aerosol layer (also known as haze) \citep{Waite2007}. 
Finally, the aerosols reach the surface, contributing to the sand dune-like topography seen in the equatorial region \citep{Lopes2016}.

The surface temperature of Titan is about 94 K at the equator \citep{Fulchignoni2005} and 91-92 K at the poles \citep{Jennings2009}. 
What makes Titan particularly interesting is that this temperature is close to the triple points of CH$_{4}$ and C$_{2}$H$_{6}$, which are minor atmospheric components (90.67 K and 90.35 K, respectively). On Earth, H$_{2}$O exists in various phases--gas, liquid, and solid due to the surface temperature being near H$_{2}$O's triple point. Similarly, Titan has what can be called a methane cycle, dominated by CH$_{4}$. 
The driving force behind this cycle, as on Earth, is solar energy. 
Titan is the only known object other than Earth on which stable liquid exists. The Cassini spacecraft, which began observing Titan in 2004, confirmed that large clouds of CH$_{4}$ formed in the high latitudes of Titan's southern hemisphere, followed by the appearance of lake-like features on the surface \citep{Turtle2011a}. 
At that time, it was summer in Titan's southern hemisphere, indicating that CH$_{4}$ evaporation, cloud formation, and rainfall driven by solar energy were actively occurring. In 2005, the Huygens probe released from the Cassini spacecraft and landed on Titan, marking the first landing on an icy body. Huygens took images during its descent through the atmosphere and revealed dendritic patterns on the surface, likely formed by liquid CH$_{4}$ flowing into river channels after precipitation (Figure 8). Saturn has an equatorial tilt of about 25 degrees and completes an orbit around the Sun in roughly 30 years. As the Cassini mission continued its observations, by around 2010, CH$_{4}$ clouds were observed at mid-to-low latitudes \citep{Turtle2011b}. 
Besides this vertical methane circulation, Titan also experiences north-south methane circulation between its summer and winter poles. In the Arctic, where it was winter in the mid-2000s when the Cassini arrived, several lakes were identified, some as large as North America's Great Lakes \citep{Stofan2007}. 
In the south pole region, where it was summer at the time, the levels of these lakes were observed to gradually decrease \citep{Hayes2011}. 
As mentioned, CH$_{4}$ in Titan's atmosphere is broken down by photochemical reactions in the upper atmosphere, producing polymeric organic materials such as aerosols, leading to a gradual depletion of CH$_{4}$ over time. This rate of CH$_{4}$ loss suggests that the gas could be depleted from Titan's atmosphere in 10 to 30 million years \citep{Atreya2010}. 
It is believed that CH$_{4}$ is continuously supplied from Titan's interior through processes such as cryovolcanism. However, no definitive evidence of cryovolcanic features has been observed so far. Meanwhile, Cassini measured the tidal response of Titan and found it to be significantly greater than expected for a fully solid body, suggesting the existence of a subsurface ocean \citep{Goossens2024}.

Titan's atmosphere and surface provide a potential environment for a surface biosphere that relies on solar energy, similar to Earth. While liquid H$_{2}$O works as the solvent for chemical reactions on Earth, hydrocarbons could play this role on Titan. Although the detailed understanding of this environment remains unclear, the presence of complex organic molecules, solar energy, and hydrocarbons with its low melting point suggest the possibility of chemical evolution. Further research is required to understand the potential for life in such an environment, and Titan offers a novel perspective, expanding our understanding of life beyond the Earth-centered view that depends on liquid H$_{2}$O as a solvent. Moreover, the possibility of a subsurface ocean suggests that Titan could host not only an Earth-like surface biosphere but also a deep biosphere.
\begin{figure}[htbp]
  \centering
  \includegraphics[width=1.0\linewidth]{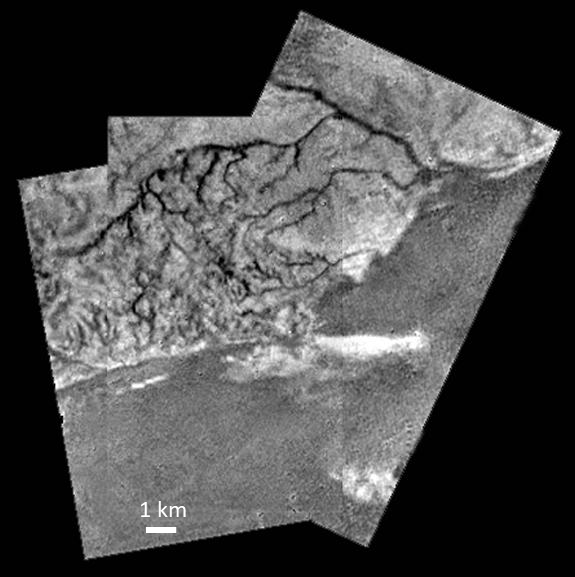}
  \caption{This mosaic of three frames taken by the Huygens Descent Imager/ Spectral Radiometer (DISR) instrument (PIA07236).}
  \label{fig:figure8}
\end{figure}
%
\subsection{Iapetus}

Iapetus is the third-largest moon of Saturn with an estimated mean density of 1,090 kg/m$^{3}$ suggesting made up mostly of ice. Most striking feature is a distinctive difference in surface coloration between its bright ($\sim$60\% in visible) leading hemisphere and its dark ($\sim$5\%) trailing hemisphere \citep{Hendrix2008}. 
Due to Iapetus's slow rotation, the surface temperature during the day attains nearly 130 K in the dark regions, owing to their high thermal absorption, while it remains around 110 K in the bright regions \citep{Spencer2010}. 
Such temperature difference causes ice to sublimate preferentially from the dark regions and deposit onto the bright, colder regions and high-latitude areas. As a result, over geological timescales, a positive runaway feedback occurs, wherein the dark regions become darker and the bright regions become brighter, enhancing the albedo contrast \citep{Spencer2010,Kimura2011}.
One hypothesis proposes that the dark material on Iapetus may have originated from other moons, specifically from the outer moon Phoebe, which has a very low albedo \citep{Verbiscer2009}. 
The composition of the dark material remains unknown, however, some experimental studies and spectral analyses suggest that a mixture of tholin, carbon dioxide, aromatic hydrocarbons, and other non-water compounds could be potential candidates \citep{Filacchione2007,Cruikshank2008,DalleOre2012}. 
Traces of carbon dioxide ice and possibly ammonia ice have also been detected, particularly in darker areas \citep{Clark2005,Palmer2011}.

Current overall shape exhibits equatorial flattening consistent with hydrostatic equilibrium if its spin period were about 16 hours. 
However, current period is 79 days, implying that the current shape was likely established early in its history and frozen by formation of the thick crust, then its spin gradually slowed due to tidal dissipation. The equatorial ridge runs about 1,300 km long, 20 km wide and 20 km high \citep{Porco2005} (Figure 7). 
The ridge system on Iapetus is heavily cratered, indicating it is ancient. Critical points for this ridge are, why it aligns exactly on the equator and why it is found only on Iapetus so far. Several hypotheses for the ridge formation have been suggested such as the building the ridge and despinning of Iapetus through an impact-generated disk and sub-satellite \citep{Levison2011,Dombard2012}, and the convective overturn of a high-density core enclosed by the ice layer as a result of melting, gravitational differentiation and freezing \citep{Czechowski2013}. 
However, the formation process of the ridge remains unclear. 
Iapetus' surface is primarily composed of water ice, especially on the bright trail-ing hemisphere. The leading hemisphere, which appears dark, contains carbonaceous material, likely including hydrocarbons and tholin-like organics. Traces of carbon dioxide ice and possibly ammonia ice have also been detected, particularly in darker areas. The contrast between the two hemispheres is a result of both external and internal processes, including material deposition from Phoebe and endogenic resurfacing.
%
\subsection{Hyperion and Phoebe}

Hyperion revolves around Saturn every 21.3 days in the orbit between Titan and Iapetus. Its (180\,$\times$\,133\,$\times$\,83 km) spin axis has a chaotic orientation in time \citep{Black1995}: rotating nearly around the long axis, at $\sim$72$^{\circ}$-75$^{\circ}$ per day \citep{Thomas2007}. 
A mean density of 538.6\,$\pm$\,48 kg\,m$^{-3}$ indicates a porosity of $\sim$46\% if Hyperion is primarily water ice \citep{Jacobson2022}. 
Hyperion has a unique sponge-like appearance. Surface is dominated by impact craters displaying deep, sharp-edged morphology, and showing little evidence of ejecta. 
Experiments and scaling theory suggest that impacts into high-porosity targets generate less ejecta mass when craters are primarily formed through compression rather than excavation \citep{Thomas2007}. 
Infrared spectroscopy by the Cassini spacecraft found the absorption band of the CO$_{2}$ molecule, and the band is shifted in wavelength from the nominal 4.27\,$\mu$m to 4.25\,$\mu$m, indicating that CO$_{2}$ on Hyperion is chemically or physically complexed with some other material. 
Also, it founds spectral similarities in the low-albedo material on both Hyperion and Iapetus, indicating widespread complex organic solid material-and possible aromatics.

Phoebe is the most massive irregular satellite of Saturnina system (110\,$\times$\,109\,$\times$\,102 km in radial dimensions) \citep{Thomas2010}, rotating in a retrograde orbit 10 times farther from Saturn than Titan. 
In contrast to most Saturnian moons, Phoebe's surface albedo is much lower ($\sim$0.10) \citep{Grav2015}. 
In contrast, the close-up view taken by the Cassini spacecraft reveals that impact craters show a large local variation in brightness, and that bright material appears to be exposed on crater slopes and by mass wasting of steep scarps. It indicates that the presence of bright ice below a relatively thin (from a few meters to a few 100s meters thick) blanket of dark surface deposits \citep{Porco2005}. 
Spectroscopic observations of its surface have revealed the presence of carbon dioxide. Also, because of its relatively large bulk density of 1640 kg/m$^{3}$, which is relatively large among Saturn's moons, Phoebe is thought to be a captured centaur. In planetary science, the term "centaur" refers to a small body that originated in the outer solar system, typically between Jupiter and Neptune, and later migrated inward. 
Phoebe is thought to have been one such object that was subsequently captured by Saturn's gravity \citep{Johnson2005}. 
The material ejected from the surface by micrometeoroid impacts is considered to spread around the Phoebe's orbit, forming a faint ring detected by the Spitzer Space Telescope \citep{Verbiscer2009}. 
This dust ring extends from around the Iapetus's orbit to  beyond Phoebe's, with a thickness estimated at approximately 40 times Saturn's radius. This dust ring may be the source of the dark material on the Iapetus's leading hemisphere.
%
\subsection{Miranda, Ariel, Umbriel, Titania and Oberon}

Uranus has 28 confirmed moons (as of August 2025). 
The surfaces of all these moons are covered with ice, and five of them--Miranda, Ariel, Umbriel, Titania, and Oberon, listed in order from closest to Uranus--have radii of several hundred kilometers. All these moons synchronously orbit close to Uranus's equatorial plane in nearly circular orbits, with orbital periods ranging from 1.4 days (Miranda) to 13.5 days (Oberon). These moons orbit the Sun in a manner similar to Uranus, with their axes tilted almost sideways relative to their rotation. As a result, their northern and southern hemispheres face either directly towards or directly away from the Sun during the solstices. 

Their bulk densities range from about 1,200 to 1,700 kg/m$^{3}$, with higher densities indicating a greater proportion of rock relative to ice. However, since detailed close-flyby explorations have not been conducted on all these moons, their moments of inertia and the degree of internal differentiation remain unknown. 
On the other hand, these moons display a wide range of geological features on their surfaces. While some, Miranda, Ariel, and Titania feature large rift valleys and flow topography, which are believed to be remnants of significant past geological activity, others, Umbriel and Oberon, lack notable geological features, and their surfaces are heavily cratered. This diversity highlights the complex and varied history of the Uranian system. Because the Voyager 2 spacecraft provided the first and only in-situ reconnaissance during its flyby to Uranus, geological and compositional information for their moon is highly limited. They can preserve only a few tens of kilometers of ocean until present if the oceans are maintained by antifreeze, such as ammonia and chlorides \citep{CastilloRogez2023}.

Miranda is the smallest and innermost of Uranus's major moons, with a radius of approximately 236 km. With a density of about 1,200 kg/m$^{3}$, Miranda is primarily com-posed of water ice mixed with some rock. Although a detailed compositional map does not exist, ammonia has been identified on its surface \citep{Bauer2002}.
Miranda's surface is divided into ancient cratered plains, estimated to be approximately 3.9-4.5 billion years old, and younger resurfaced areas. The latter features circular to rectangular outlines and complex morphologies, known as coronae, which are thought to be a few hundred million years old \citep{Plescia1988}. 
It has one of the most intriguing surfaces in the Solar System, characterized by massive cliffs, deep fault canyons, and patchwork terrains, all indicative of intense geological activity. The surface composition appears to be a mixture of ice and silicate rock, and these varied features suggest a complex history in-volving tectonic activity and possibly cryovolcanism \citep{Schenk1991}. 
The distribution of coronae on Miranda may reflect the dynamic processes within its interior. Its resemblance to the major resurfacing zones on Enceladus, as well as the distribution of relaxed craters on Dione, suggests that some mid-sized moons exhibit a characteristic bi-axial mode in which heat and mantle materials are redistributed \citep{Schenk2020}.

Ariel, with 579 km in radius, is one of the brightest moons of Uranus, the Bond albedo is about 23\%\citep{Karkoschka2001}. 
Ariel's density, approximately 1,517 kg/m$^{3}$, indicates a composition of roughly half and half on ice and rock in mass. 
There is no observational evidence for subsurface oceans at present, because the Voyager 2 did not get close enough to detect any signatures. However, oceans were possible in the past based on the speculations from resurfaced signatures on Ariel. The surface can be divided into three units, cratered terrain, ridged terrain and plains \citep{Plescia1987}. 
Cratered terrain is the most widespread surface unit, featuring impact craters on a rolling landscape with narrow ridges and grabens, known as chasmata which is typically 15-50 km wide, intersecting it. 
The terrain has a blotchy appearance due to variations in albedo, and some are-as show muted topography, which is not caused by lighting conditions but may indicate degradation. Ridged terrain, forming polygonal patterns that enclose the cratered terrain, consists of east- or northeast-trending bands with parallel ridges and troughs. These bands extend for hundreds of kilometers with 25-70 km wide  and are characterized by a high density of faults, suggesting formation under tensional stress. The connection between the ridged terrain and the chasmata remains uncertain. 
Identical crater distribution to that for the cratered terrain indicates similar surface age to that terrain \citep{Plescia1987}. Plains are smoother, less cratered areas in low regions like graben floors. 
They exhibit lobate margins, suggesting that plains materials were emplaced as flows. In chasmata, plains have central troughs with brighter, smoother material indicating lateral flow. Troughs also ap-pear at the plains' edges, but their role as vents is unclear. Plains display variable crater frequencies, and a few of younger areas appear to be less than 100 Myrs old, suggesting that Ariel may still be geologically active \citep{Plescia1987}. Spectroscopic observations us-ing the ground-based telescope have found CO$_{2}$ \citep{Grundy2006} and ammonia \citep{Cartwright2020} on Ariel.

Umbriel, in contrast to its neighboring moons, is darker and older. A radius is approximately 585 km and bulk density is about 1539 kg/m$^{3}$, indicating a composition that includes both rock and water ice, similar to that of Ariel. The heavily cratered surface shows fewer signs of tectonic or cryovolcanic activity compared to Ariel and Miranda, suggesting that Umbriel's interior has likely been less geologically active. The surface, which has the lowest bond albedo ($\sim$0.10) among the Uranian moons \citep{Karkoschka2001}. 
The absorption bands of water ice are stronger on the leading hemisphere than on the trailing hemisphere, likely due to bombardment by charged particles from Uranus's magnetosphere \citep{Grundy2006}. CO$_{2}$ ice has been detected on the trailing hemisphere \citep{Grundy2006,Cartwright2015}, and it migrates to lower latitudes over geo-logically short timescales. 
Near the equator, Wunda Crater, which has a diameter of ap-proximately 130 km, features a large ring of bright material on its floor. This material is likely a deposit of pure carbon dioxide ice formed when radiolytically produced CO$_{2}$ migrated across Umbriel's surface and became trapped in the relatively cold environment of Wunda \citep{Sori2017}.

Titania, the largest of Uranus's moons, has a radius of approximately 789 km. With a density of about 1,680 kg/m$^{3}$, Titania consists of roughly 60\% rocky material by mass, with the remainder being water ice. Observations reveal cratered plains and signs of tectonic activity, including faults, cliffs, and potential ice flows. 
These features suggest geological processes likely driven by the expansion or contraction of its crust due to thermal changes over time \citep{Smith1986}. 
The surface is abundant with impact craters and is intersected by a network of large faults and scarps. Two parallel scarps form grabens, also sometimes referred to as canyons (chasmata). The grabens range from about 20 to 50 km in width, and the relief of the scarps is approximately 2 to 5 km \citep{Smith1986}. 
This faulting indicates a global extension of Titania's crust, which may have occurred in response to the final stages of ice freezing within the moon's interior. 
Craters on Titania that are larger than 150 km in diameter have shallow, flat floors, indicating that they have been degraded or relaxed over time.

Oberon, similar in size and bulk density to Titania, has a radius of 761 km and a density of 1,682 kg/m$^{3}$. It is the outermost of Uranus's large moons. 
However, its cratering record is quite different. The heavily cratered, dark, and ancient surface shows limited signs of geological activity, suggesting that its interior has likely remained cold and inactive. While some linear features of possible tectonic origin can be observed, their identification is inconclusive due to the low resolution of available surface images. A few bright craters with dark deposits may indicate the excavation of darker subsurface material and the formation of bright, impact-related frost or debris \citep{Smith1986}. Although the surfaces of Uranus's major moons tend to darken on the trailing hemisphere due to bombardment by Uranian magnetospheric plasma, Oberon does not exhibit this pattern \citep{Grundy2006}.
%
\subsection{Triton}

Neptune has 16 moons (as of September 2024), with Triton being by far the largest, having a radius of 1,353 km. Its bulk density of 2,061 kg/m$^{3}$ indicates a composition of approximately 30\,-\,45\% water ice by mass (Figure 1). 
However, the degree of internal differentiation remains unclear due to the lack of gravity field measurements and the unknown moment of inertia. 
Triton is thought to have originated as a trans-Neptunian object, which was later captured by Neptune, as many of the other smaller moons having large elliptical orbits or retrograde orbits.
Triton's surface is covered by frozen nitrogen and shows very few impact craters, as observed by the Voyager 2 spacecraft during its close flyby in 1989, when it observed part of the southern hemisphere. 
The estimated surface age is less than 100 million years, indicating that Triton's surface is significantly young. 
The surface is dominated by regions with mountain ranges and canyons forming a mesh-like pattern, resembling a "cantaloupe," along with vast areas that are remarkably flat with minimal topographic variation. 
Cryovolcanism, the eruption of substances like liquid nitrogen and methane, has been ob-served in some regions. 
This cryovolcanism is thought to be sustained mainly by seasonal variations in solar energy input, and previous studies have also suggested that obliquity tides may contribute as a modest additional source of internal heating \citep{Nimmo2015}. 
As a result, Triton is considered one of the most geologically active worlds in the Solar System.
Triton's atmosphere is primarily composed of nitrogen, with a surface pressure of only about 1.4 Pa  \citep{McKinnon2014}. 
Methane and carbon monoxide are present as minor components, and a haze, largely composed of hydrocarbons and nitriles produced by the photochemical breakdown of methane, is found throughout the troposphere. 
Nitrogen ice clouds are also present, forming between 1 and 3 km above the surface. Since the Voyager 2 flyby in 1989, Earth-based telescopic observations of stellar occultations have monitored Triton's atmosphere. Events in 1995 and 1997 showed a significant in-crease in atmospheric pressure relative to the Voyager 2 measurements. 
However, observation of the event in 2017 indicated that the pressure had returned to 1989 levels. 
A 2022 stellar occultation analysis derived a surface pressure of $\sim$1.45\,Pa, consistent with 2017 values and Voyager-era estimates, challenging models that predict a steady pressure decline. 
This stability hints at complex N$_{2}$ volatile cycles and surface-atmosphere feedbacks on Triton \citep{Sicardy2024}.
%
\subsection{Icy Dwarf Planets}
All four dwarf planets, except Ceres, which is located in the main asteroid belt, are situated in the outer solar system beyond Neptune's orbit. Among them, Pluto is the largest. Pluto has a radius of 1,188 km and an orbital semi-major axis of 39.5 au (the astronomical unit, which is a unit of length defined to be $\sim$149,600,000 km) with an orbital inclination of 17.2 degree. The New Horizons spacecraft flew by Pluto in July 2015 and performed surface imaging, radio science investigation, spectroscopy and other experiments \citep{Stern2018}. 
With an average density of approximately 1,854 kg/m$^{3}$, Pluto's interior is believed to be composed of both ice and rock. Its very slight oblateness (<0.6\%) indicates that Pluto has lost the larger fossil bulge expected for a cold, rigid body, implying viscous relaxation of the ice shell consistent with a warm, deformable interior and the possible presence of a subsurface ocean (Figure 3) \citep{Nimmo2017}. 
Additionally, a geophysical analysis of the location of a major geological feature on Pluto, Sputnik Planitia (the western lobe of Tombaugh Regio), supports the existence of a subsurface ocean \citep{Nimmo2016,Kamata2019,Kimura2020}. 
Moreover, strong spectral signatures of H$_{2}$O ice detected on extensional faults \citep{Grundy2016} imply that Pluto may have undergone global expansion, possibly due to the freezing of a thick subsurface ocean in the past. 
The ice on Pluto's surface is not solely H$_{2}$O; it also contains significant amounts of solid methane and nitrogen, which are unevenly distributed. 
In the equatorial region, there is a heart-shaped area characterized by smooth, white terrain, most of which is covered in nitrogen ice. 
This region shows almost no impact craters and instead displays features resembling the flow of highly plastic material, along with cell-like or honeycomb structures (Figure 9). 
Pluto's surface is extremely cold, with an average temperature of about 44 K. Pluto also has a thin atmosphere composed of nitrogen, methane, and carbon monoxide. During the Pluto flyby of the New Horizons, 
Charon, the largest satellite of Pluto, was also observed in detail, revealing a surface dominated by H$_{2}$O ice, a large equatorial trough system, and extensive tectonic features indicative of past internal activity \citep{Moore2016,Beyer2017}.

Haumea is a dwarf planet located beyond Neptune's orbit, with a semi-major axis of 43.1 AU, an inclination of 28 degree and an orbital period of 283 Earth's year. Ground-based telescopic spectrometry has shown strong crystalline water ice features on its surface, with temperatures below 50 K \citep{Trujillo2007}. 
Homogeneous water ice, consisting of a 1:1 mixture of crystalline and amorphous ice, covers about 80\% of the surface. There is no methane detected, and only a few percent of organic materials are present. 
Observations of a stellar occultation revealed a strongly elongated shape, with a long axis of 1,960 km and a short axis of 996 km \citep{Dunham2019}. 
The light curve from telescopic observations indicates an extremely short spin period of 3.9 hours \citep{SantosSanz2017}, suggesting that such a rapid spin would cause the planet to distort into a triaxial ellipsoid.

Makemake follows an orbit similar to that of Haumea, with a semi-major axis of 45.4 AU, an inclination of 29 degrees, and an orbital period of 306 Earth years. Based on photometric observations, the spin period was estimated to be about 22.8 hours \citep{Hromakina2019}. 
The near-infrared spectrum shows a weak absorption signature of methane and the possible presence of ethane and tholins due to the photolysis of methane by solar radiation \citep{Brown2007}. 
Stellar occultation observations revealed that Makemake has no global atmosphere, with an upper limit of around 10 nanobars (10$^{8}$ times less than Earth's atmosphere) at the surface \citep{Ortiz2012}.

Eris is the furthest and most massive dwarf planet in the Solar System, with an orbital semi-major axis of 67.9 AU, an eccentricity of 0.436, and an inclination of 44.0 degrees. 
Eris has an orbital period of 559 Earth years, and the distances of the perihelion and aphelion are 38.0 and 97.5 AU, respectively. 
Near-infrared spectroscopy revealed the presence of deuterated methane ice on its surface, and its low deuterium abundance suggests that Eris's methane is not primordial but may have undergone geochemical process-es. 
In addition, nitrogen ice has been identified, with an inferred volume of one-third that of methane \citep{Grundy2024}. 
Although sublimation models suggest that the surface nitrogen could create an atmosphere \citep{Hofgartner2019}, this has not been directly confirmed by observations.

\begin{figure}[htbp]
  \centering
  \includegraphics[width=1.0\linewidth]{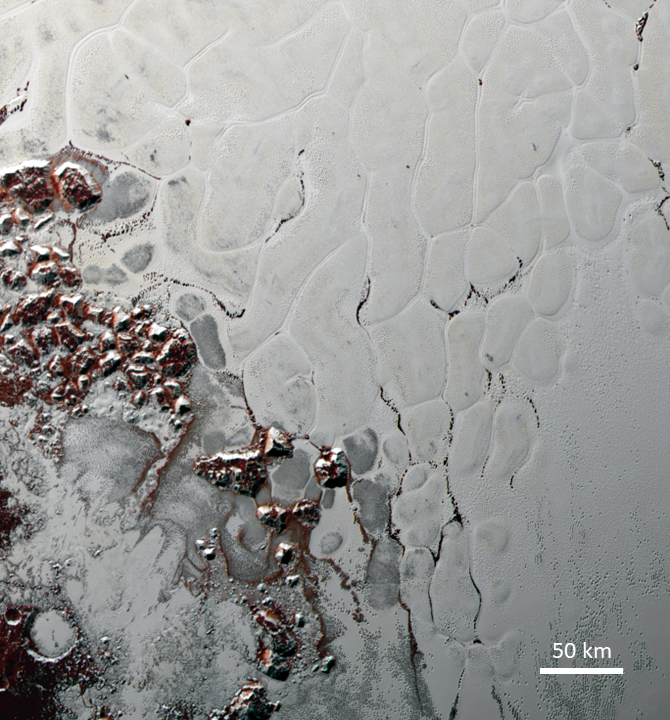}
  \caption{Western region of Sputnik Planitia on Pluto provided from the New Horizons Ralph/Multispectral Visible Imaging Camera (MVIC) (PIA20726).}
  \label{fig:figure9}
\end{figure}
%
\section{Questions and Future Prospects}
\subsection{Intermittent Activity and Plume Mechanisms}

Plume activity has been reported intermittently at Europa and persistently at Enceladus, but the physical mechanisms driving such activity are still debated. For Europa, Hubble Space Telescope observations suggested episodic water vapor plumes, yet follow-up studies including JWST have failed to confirm them consistently, leading to the view that the activity is highly variable or spatially confined. Enceladus, in contrast, exhibits a sustained plume originating from its south polar terrain, although temporal variability in flux has also been proposed. These contrasting cases highlight the need for comparative studies. Future campaigns combining JWST spectroscopy, long-term monitoring by ground-based telescopes, and targeted in-situ flybys by Europa Clipper and JUICE are required to constrain eruption frequency, mass flux, and compositional variability. Understanding the physical triggers, such as tidal stresses, cryovolcanic conduits in ice shells or overpressure in subsurface oceans will be critical to linking plume activity with the subsurface oceans that may sustain habitable environments.
\subsection{Redox Budgets and Ocean Delivery}

Habitability assessments for icy bodies depend critically on the availability of chemical energy. Juno's recent detection of O$_{2}^{+}$ and H$_{2}^{+}$ ions at Europa now provides quantitative estimates of surface oxidant production. 
However, the key unknown remains how these oxidants are transported through tens of kilometers of ice shell into the subsurface ocean. 
Several mechanisms have been proposed, including subduction of surface ice, melt-through processes, and transport along fractures. 
Each process implies largely different timescales and fluxes. 
Europa Clipper's gravity and magnetometer data, combined with thermal imaging and spectroscopy, should reveal how the ice shell recycles material and whether oxidants are delivered efficiently enough to sustain redox gradients for life. 
A parallel issue is the potential reduction of ocean chemistry by hydrothermal activity at the seafloor. 
The balance between oxidant input from the surface and reductant production in the interior represents a central question in evaluating Europa's biosignature potential.
\subsection{Young Ocean Worlds Beyond the "Usual Suspects"}

Traditionally, discussions of ocean worlds have focused on a limited set of major satellites, such as Europa, Enceladus, and Titan. 
These are the usual suspects, but recent works suggest that smaller and less-studied icy bodies may also host subsurface oceans, with Mimas being a striking candidate. 
If confirmed, this discovery would fundamentally expand the range of ocean worlds and force us to reconsider how widespread such environments may be. 
Other small icy bodies, long regarded as inactive, could also harbor recently formed or transient oceans. 
This motivates a broader re-investigation of icy bodies, using dynamical models, tidal heating studies, and reanalysis of existing datasets. 
Expanding the search beyond the usual suspects could reveal a larger diversity of oceanic environments, with profound implications for habitability across the solar system.
\subsection{Interior Architecture of the Icy Galilean Moons}

The possibility of subsurface oceans within the Galilean satellites has been a central theme in icy body research. 
However, the internal structures of the Galilean icy satellites remain a subject of considerable uncertainty, and the degree of confidence differs strongly among them. 
Although current interior models must balance competing constraints from gravity data, induced magnetic fields, and surface geology, significant degeneracies persist. 
For Europa, multiple lines of evidence provide strong support for a global ocean beneath its ice shell. For Ganymede, the evidence is more ambiguous. 
Proposed signatures of magnetic induction are consistent with a conductive ocean, although separating a putative induced signal from the strong intrinsic dipole field remains unresolved. 
If an ocean does exist, theoretical models allow for the possibility of multiple stacked liquid layers, separated by high-pressure ice phases. 
Such a configuration would raise questions about long-term stability, heat transfer, and the potential for chemical communication with the surface. 
Callisto provides a contrasting example. 
Although its bulk size and density are similar to those of Ganymede, its ancient, heavily cratered, and relatively unmodified surface indicates a limited degree of internal activity. 
The high moment of inertia factor has been interpreted as evidence for only partial differentiation, although its precise value remains debated.  
Although magnetic observations have been interpreted as consistent with induction in a saline liquid layer, the depth, extent, and even the existence of such an ocean remain uncertain. 
Current models of Ganymede and Callisto must balance competing constraints from gravity data, magnetic signatures, and surface geology, but significant degeneracies persist. 
Forthcoming data from JUICE and Europa Clipper, gravity harmonics, magnetic induction, and rotational state, should clarify the architecture of these satellites and test competing hypotheses. 
Improved constraints will be essential for assessing the diversity of possible ocean worlds, and for placing Europa's comparatively well-supported case within a broader comparative framework.
\subsection{Volatile Cycles in Distant Icy Worlds}

Triton remains one of the most enigmatic icy bodies. 
Observations from stellar occultations in 2017 and 2022 indicate that its tenuous nitrogen atmosphere has remained nearly stable for several decades, contrary to models predicting a steady decline in surface pressure.
Such stability suggests that complex feedback processes between surface ices, seasonal insolation, and atmospheric escape operate on Triton. 
These processes are not only relevant for understanding Triton itself but also for comparative studies with Pluto and other Kuiper Belt objects that share similar volatile cycles. 
However, the current dataset is limited to a handful of occultations. 
Multi-chord occultation campaigns, thermal-infrared mapping, and long-term monitoring of surface albedo patterns will be required to con-strain volatile transport models. 
These studies are also directly relevant to understanding the evolution of volatile retention and atmospheric dynamics on other distant icy worlds in the outer solar system.
\subsection{Mission Outlook and Strategic Priorities}

The coming next decade offers a remarkable set of opportunities for in-situ exploration of icy bodies. 
Europa Clipper will provide multi-instrument observations of Europa's surface, atmosphere, and interior from 2030 onward, yielding transformative in-sights into its ocean, ice shell, and habitability potential. 
With similar timing, JUICE mission, led by European Space Agency (ESA), will explore Jupiter's icy Galilean moons through multiple flybys from 2031, eventually entering orbit around Ganymede. 
Lucy spacecraft was launched in 2021 to explore the Jupiter Trojans, which are thought to be remnant embryos of icy bodies, with an expected arrival at the Trojans in 2027. 
In 2028, Dragonfly, a rotorcraft mission to Titan, is planned for launch, with objectives including the study of surface composition, atmospheric analysis, and seismic measurement at multiple sites on Titan's surface. 
Looking further ahead, NASA has selected the Uranus Orbiter and Probe as its next flag-ship-class mission in the highest priority, expected to bring a big leap for our understanding of the Uranian system, one of the least understood among the icy bodies.
Laboratory experiments on ice rheology, volatile chemistry, and radiation processing, together with improved theoretical models, will be essential to interpret spacecraft data. 
Strategic coordination across observational platforms and disciplines will maximize the scientific gain from these missions and will drive a new era of discovery. 
The overall outcome of these initiatives is expected to integrate discoveries such as subsurface oceans, persistent atmospheres, and unexpected surface processes into a coherent framework for understanding the formation, evolution, and potential habitability of icy bodies in the solar system. 
And finally it will be central to understanding their potential as habitable environments and their broader role in planetary system evolution.

%
\section{Conclusion}

This review has synthesized current knowledge of icy bodies based on observations, in-situ exploration, and theoretical modeling. Recent advances have provided new constraints on atmospheric composition, internal structure, and surface activity through data from Juno, JWST, and stellar occultations. Decades of spacecraft investigations, including Voyager 1 and 2, Galileo, Cassini-Huygens, New Horizons, and Juno, together with ground- and space-based telescopic observations, have greatly expanded our under-standing of the diversity of icy bodies in the solar system. However, these efforts have also revealed fundamental gaps.

Progress remains limited by several factors. First, the chemistry of icy materials is poorly constrained because no physical samples have been returned. For silicate bodies, we have only obtained samples from the Moon and from asteroids, and no comparable material has been collected from icy bodies. Second, the geological history of icy surfaces is difficult to reconstruct owing to uncertainties in the size distribution of small bodies in the outer solar system and the incompletely constrained impact histories. Third, the coupling between internal and orbital evolution remains poorly quantified because the influence of tidal interactions between planets and their moons is difficult to evaluate.

Although these limitations cannot be resolved immediately, a broad suite of missions is now in progress or planned. JUICE and Europa Clipper will investigate the Galilean satellites with unprecedented detail. Lucy will explore the Jupiter Trojans, which may preserve records of primordial icy bodies. Dragonfly will assess Titan's surface and atmosphere, and the planned Uranus Orbiter and Probe will provide the first comprehensive exploration of the Uranian system. These efforts will be supported by advances in laboratory experiments on ice properties, chemical processes, and radiation effects, together with increasingly sophisticated numerical models. 

The next two decades are expected to yield major advances in our understanding of icy bodies. Discoveries of young oceans in small satellites, the persistence of tenuous atmospheres in distant objects, and the likely detection of extrasolar analogues all point to the diversity and dynamism of these systems. Addressing the outstanding questions summarized here will be central to assessing the potential habitability of icy environments and clarifying their role in the formation and evolution of planetary systems.
%
\section{Suggested Reading}
Grasset, O., Blanc, M., Coustenis, A., Durham, W. B., Hussmann, H., Pappalardo, R. T., Turrini, D. (Eds.). (2010). Satellites of the Outer Solar System. Springer.

Gudipati. M. S., Castillo-Rogez, J. (Eds.). (2013). The Science of Solar System Ices. Springer.

Yamagishi, A., Kakegawa, T., Usui, T. (Eds.). (2019). Astrobiology. Springer.

\section*{Acknowledgments}
The author appreciates the reviewers for their constructive comments and suggestions, which helped improve the clarity of this article.

\bibliographystyle{elsarticle-harv} 
\bibliography{ref}

\end{document}